\documentclass[12pt]{iopart}

\usepackage[utf8]{inputenc}
\usepackage{hyperref}
\hypersetup{  
colorlinks=true,
citecolor=blue} 
\usepackage{iopams} 
\usepackage{bm}
\usepackage{graphicx}

\newcommand{\f}[2]{{\frac{#1}{#2}}}
\newcommand{\text}{\mathrm}
                      
\newcommand{\I}{1}
\newcommand{\II}{2}

\newcommand{\X}{{\scriptscriptstyle{X}}}
\newcommand{\Y}{{\scriptscriptstyle{Y}}}
\newcommand{\exy}{{(x,y)}}
\newcommand{\eyx}{{(y,x)}}

\newcommand{\Jmoyi}[1]{J_{#1}}
\newcommand{\Jstochi}[1]{j_{#1}}

\newcommand{\Jmoy}{\bm{J}}
\newcommand{\Jstoch}{\bm{j}}

\renewcommand{\geq}{\geqslant}
\renewcommand{\leq}{\leqslant}

\begin{document}

\title{Degree of coupling and efficiency of energy converters far-from-equilibrium}
\author{Hadrien VROYLANDT$^1$, David LACOSTE$^2$, Gatien VERLEY$^1$}
\address{$^1$ Laboratoire de Physique Théorique (UMR8627), CNRS, Univ. Paris-Sud, Université Paris-Saclay, 91405 Orsay, France}
\address{$^2$ Laboratoire de Physico-Chimie Th\'eorique - UMR CNRS Gulliver 7083, PSL Research University, ESPCI
  10 rue Vauquelin, F-75231 Paris, France}


\begin{abstract}
In this paper, we introduce a real symmetric and positive semi-definite matrix, 
which we call the non-equilibrium conductance matrix, 
and which generalizes the Onsager response matrix for a system in a non-equilibrium
stationary state. We then express the thermodynamic efficiency 
in terms of the coefficients of this matrix using a parametrization 
similar to the one used near equilibrium. This framework, then valid
arbitrarily far from equilibrium allows to set bounds on the thermodynamic efficiency 
by a universal function depending only on the degree of coupling between input and output currents. 
It also leads to new general power-efficiency trade-offs valid for macroscopic
machines that are compared to trade-offs previously obtained from uncertainty relations. We illustrate our results on an unicycle heat to heat 
converter and on a discrete model of molecular motor.
\end{abstract}

\section*{Introduction}
\label{sec:Intro}

The second law of thermodynamics prevents the thermodynamic efficiency 
of energy converters to exceed the reversible efficiency \cite{Callen1985_vol}, thus ruling out perpetual motion.
The energy converters operating close to reversible efficiency have been widely studied
\cite{Kedem1965_vol61,Esposito2009_vol85,Polettini2015_vol114,Polettini2017_vol118}.
Historically, these questions were first addressed 
within the framework of weakly irreversible thermodynamics developed by Onsager for purely resistive systems 
\cite{Onsager1931_vol37, Onsager1931_vol38}, i.e. systems with fluxes and affinities instantaneously related. This theory assumes 
local equilibrium and expresses physical currents (e.g. energy currents, matter currents) as non-linear functions of the affinities 
and local intensive parameters \cite{Callen1985_vol}.  

In the linear response regime near equilibrium, currents become linear function of 
the affinities, which defines the so-called Onsager response matrix.
The framework based on this response matrix has been very successful to describe thermoelectric
 effects \cite{Wood1988_vol51, Domenicali1954_vol26}, to determine 
the degree of coupling between influx and outflux \cite{Kedem1965_vol61, Caplan1966_vol10, Book_Hill1989},
 or to predict the efficiency at maximum power \cite{VandenBroeck2005_vol95, Esposito2009_vol102}.
This framework can also be extended to cover mesoscopic and nanoscale systems \cite{Benenti2017_vol694}. 
A key result of the response matrix framework is Onsager's reciprocity relations which can be deduced 
from a more general symmetry property called fluctuation theorems 
\cite{Andrieux2004_vol121,Andrieux2007_vol, Seifert2012_vol75}. 
Previous attempts to generalize the notion of Onsager matrix to non-equilibrium stationary states 
lead to non-symmetric Onsager matrices, 
so that many properties were lost for that reason~\cite{Verley2011_vol93}.

In this paper, building on the work of Polettini \etal\cite{Polettini2016_vol94, Polettini2016_vol94a}, we introduce precisely a non-equilibrium conductance matrix that is symmetric just as the Onsager response matrix, but whose coefficients are now functions of the affinities. Intuitively, such a conductance matrix should exist at the macroscopic scale, because it can be constructed by associating conductances between every pair of states 
from the microscopic scale up to the macroscopic scale. 
Naturally, the question whether a symmetric matrix can be constructed in this way 
even when the system is in a non-equilibrium stationary state requires a more careful analysis.

For this reason, we assume in a first step that such a non-equilibrium conductance matrix 
can be constructed disregarding the issue of possible non-unicity of this matrix. 
We then show that a parametrization of the thermodynamic efficiency 
introduced by Kedem and  Caplan \cite{Kedem1965_vol61} for machines near equilibrium still applies to 
a machine operating far from equilibrium. 
This parametrization involves the degree of coupling $\xi$ between the influx and outflux, which together 
with the functions $\Pi$ and $\varphi$ characterize the dissipation and therefore also the efficiency of a machine. Using this parametrization which is linked to the chosen non-equilibrium conductance matrix, we show that the efficiency  
admits a general upper bound valid arbitrarily far from equilibrium, which only depends on the degree of coupling. 
We also deduce from our framework power-efficiency inequalities that set bounds on the output power  
as a function of the machine efficiency.

In a second step, we construct the non-equilibrium conductance matrix from 
a framework based on the Large Deviation Function (LDF) of currents. 
In the linear regime near equilibrium, this LDF has a quadratic form, 
which is related to central results of Statistical Physics such Onsager relations and the Fluctuation-Dissipation theorem. In this framework, response and current fluctuations are connected through an equality.
Near a non-equilibrium stationary state, the relation between fluctuations and response 
takes instead the form of an inequality,
which states that currents fluctuations near a non-equilibrium stationary state are always more likely than those predicted by  
linear response analysis close to this point \cite{Pietzonka2016_vol93, Gingrich2016_vol116}. 
We show that a consequence of this property is 
an inequality between the non-equilibrium conductance matrix and the matrix of covariances of currents, for a certain 
matrix order among symmetric matrices~\cite{Book_Horn1985}. 
Remarkably, our result contains various interesting power-efficiency trade-offs  \cite{Pietzonka2016_vol2016, Pietzonka2017_vol}. Hence, our approach provides a unifying framework for studying and optimizing machine performance, and illustrates the relevance of the concept 
of the non-equilibrium conductance matrix. 

To summarize, in the first section, we use standard thermodynamics to constrain the form of the non-equilibrium conductance matrix 
and then we exploit a parametrization originally developped for equilibrium systems to describe the efficiency of macroscopic
thermodynamic machines operating far from equilibrium. 
In the second section, we derive an explicit formula for this non-equilibrium conductance matrix at the stochastic level.
In the third section, we obtain various power-efficiency inequalities from that framework 
which we illustrate with two examples: a three state model 
of heat to heat converter with strongly coupled heat fluxes and a discrete model of molecular motor \cite{Lau2007_vol99,Lacoste2008_vol78}.

\section{From non-equilibrium conductance matrix to constraints on power and efficiency}

\subsection{The non-equilibrium conductance matrix}
\label{sec:prop-of-G}


Machines are systems that produce on average a current against an external force, usually called thermodynamic affinity. This is achieved by using another current generated by its own affinity. Hence, a machine generically involves two affinities $F_{\I}$ and $F_{\II}$ associated with two physical currents $\Jmoyi{\I}$ and $\Jmoyi{\II}$.
In terms of the physical currents and affinities, the mean entropy production rate can be written as
\begin{equation}
\label{sigma}
\sigma =F_\I \Jmoyi\I+F_\II \Jmoyi\II,
\end{equation}
which is the sum of two partial entropy production rates denoted by $\sigma_\X = F_{\X}\Jmoyi{\X}$. 
We focus here on the steady-state regime of the machine, where 
all quantities introduced so far are time-independent. Throughout the paper, we use $k_{B}=1$ which means that entropy production rates have the dimension of inverse time.
Physical observables, including currents $J_\X$, affinities $F_\X$, and partial entropy production rates $\sigma_\X$, 
are labeled with index $\X=1,2$. Close to equilibrium, physical currents are linear functions of the affinities:
\begin{equation}
  \label{eq:LinearRegime}
  \left( \begin{array}{c}
    \Jmoyi{\I} \\ \Jmoyi{\II}
  \end{array} \right) =   \left( \begin{array}{cc}
    L_{\I,\I} & L_{\I,\II} \\
    L_{\II,\I} & L_{\II,\II} \\
  \end{array} \right)
    \left( \begin{array}{c}
    F_{\I} \\ F_{\II}
  \end{array} \right),
\end{equation}
where $L_{\X,\Y}$ are the components of the Onsager matrix $\bm{L}$ \cite{Onsager1931_vol37}. This matrix has real and symmetric coefficients which are independent of the affinities. 

Beyond the linear regime, the physical currents become non-linear functions of the affinities but it is not known whether the concept of Onsager matrix can still be used for systems in non-equilibrium stationary state. Let us assume for the moment that such a generalization exists, which we call the non-equilibrium conductance matrix $\bm{G}$. By similarity with the Onsager matrix, we assume a relation of the type
\begin{equation}
  \label{eq:NonLinearRegime}
  \left( \begin{array}{c}
    \Jmoyi{\I} \\ \Jmoyi{\II}
  \end{array} \right) =   \left( \begin{array}{cc}
    G_{\I,\I} & G_{\I,\II} \\
    G_{\II,\I} & G_{\II,\II} \\
  \end{array} \right)
    \left( \begin{array}{c}
    F_{\I} \\ F_{\II}
  \end{array} \right),
\end{equation}
with again real and symmetric coefficients. An important difference with the previous case is that the coefficients of the matrix $\bm{G}$ are now necessarily functions of the affinities $F_{\I}$ and $F_{\II}$ unlike the constant coefficients of the Onsager matrix $\bm{L}$. Importantly these assumptions together with Eq.~(\ref{eq:NonLinearRegime}) still do not define a unique matrix $\bm{G}$.

We now specialize to a thermodynamic machine by assuming (without loss of generality) that the first process is the driving process and the second process is the output process. Hence the partial entropy production rate of the first process verifies $\sigma_{\I} \geq 0$ while $\sigma_{\II} \leq 0$ for the second process. The thermodynamic efficiency reads
\begin{equation}
\label{eq:efficiency}
\eta = \f{-\sigma_{\II}}{\sigma_{\I}}.
\end{equation}
The second law imposes the positivity of the total entropy production rate $\sigma=  \sigma_{\I} +  \sigma_{\II} \ge 0$, which implies $0 \leq \eta \leq 1$, where $0$ is reached when there is no output current  and $1$ is reached for a reversible operation of the machine with vanishing entropy production rate $\sigma$. 
Now, using the above properties of the non-equilibrium conductance matrix, we get for the partial entropy production rates 
\begin{eqnarray}
  \label{eq:NonEqRespMatrix}
  \sigma_{\I} & =&  \Jmoyi{\I} F_\I = G_{\I,\I} {F_{\I}}^2 + G_{\I,\II} F_{\I}F_{\II}, \\
  \sigma_{\II} & =& \Jmoyi{\II} F_\II  = G_{\II,\I} F_{\I}F_{\II} + G_{\II,\II} {F_{\II}}^2.  \label{eq:NonEqRespMatrix2}
\end{eqnarray}
We choose the affinity dependent matrix to be positive semi-definite 
to guarantee the validity of the second law for arbitrary affinities. Since $G_{1,2} = G_{2,1}$ this means:
\begin{eqnarray}
  \label{eq:def-positive}
 G_{\I,\I}G_{\II,\II} \geq {G_{\I,\II}}^2,
\end{eqnarray}
for all possible affinities. Using Eqs.~(\ref{eq:NonEqRespMatrix}--\ref{eq:NonEqRespMatrix2}) combined with the conditions $\sigma_{\I} \geq 0$ and $\sigma_{\II} \leq 0$ leads to the inequalities 
\begin{equation}
  \label{eq:contrainteEngine}
G_{\I,\I} F_{\I}^2 \geq -G_{\I,\II}F_{\I}F_{\II} \geq G_{\II,\II}F_{\II}^2 \geq 0, 
\end{equation}
which are also valid for arbitrary affinities.

The question of the existence of a non-equilibrium conductance matrix with the above properties can be resolved
by exhibiting a particular solution. It is simple to check that the matrix 
\begin{equation}
  \label{eq:GTC}
  \bm{G}_{min} \equiv \f{1}{\sigma}
  \left( \begin{array}{cc}
  \Jmoyi{\I}^2& \Jmoyi{\I}  \Jmoyi{\II} \\
            \Jmoyi{\I}  \Jmoyi{\II} & \Jmoyi{\II}^2 \\
  \end{array} \right),
\end{equation}
satisfies Eq.~(\ref{eq:NonLinearRegime}) by construction and is positive semi-definite  
because its trace is positive and its determinant is zero. The reason 
for the subscript ``min'' for this matrix, can be understood once
we introduce a matrix order for symmetric matrices, called 
Loewner partial order \cite{Book_Horn1985}. This is defined in such a way that 
$\bm{A} \geq \bm{B}$ means that $\bm{A}-\bm{B}$ is a 
positive semi-definite matrix. 
This property implies that for two symmetric $n\times n$ matrices $\bm{A}$ and $\bm{B}$:
\begin{equation}
  \label{eq:loewnerpartialorder}
  \bm{A} \geq \bm{B} \Leftrightarrow \forall \bm{x} \in  \mathbb{R}^n, \quad  \bm{x}^{\rm{T}} \cdot \bm{A} \cdot 
\bm{x} \geq \bm{x}^{\rm{T}} \cdot \bm{B} \cdot \bm{x}.
\end{equation}
Now, using Eqs.~(\ref{sigma}), (\ref{eq:NonLinearRegime}), and (\ref{eq:contrainteEngine}), one 
can show explicitly that the matrix $\bm{G}-\bm{G}_{min}$ is also a positive semi-definite matrix, because its 
trace is again positive and its determinant is zero. Therefore,
we have the general property
\begin{equation}
\bm{G} \geq \bm{G}_{min},
\label{Gmin}
\end{equation}
which justifies the name $\bm{G}_{min}$ for a matrix which represents a minimum among all conductance matrices for 
the specific matrix order defined above.

\subsection{General parametrization of the efficiency}
\label{sec:first-properties-g}

Thanks to the above properties of the non-equilibrium conductance matrix, we can introduce the functions 
\begin{equation} \label{eq:new-parameters}
\Pi \equiv G_{\I,\I}F_{\I}^2, \quad \varphi \equiv \sqrt{\f{G_{\II,\II}F_{\II}^2}{G_{\I,\I}F_{\I}^2}}, \quad \text{and}\quad \xi \equiv \f{G_{\I,\II}}{\sqrt{G_{\I,\I}G_{\II,\II}}}\, \text{sign}(F_{\I}F_{\II}).
\end{equation}
These functions are direct generalizations of the ones 
used in the close-to-equilibrium regime \cite{Kedem1965_vol61}. 
The parameter $\Pi = \Pi(F_{\I},F_{\II})$ determines the dissipation of the driving process when 
there is no output process coupled to the driving process or when there is one 
but we choose to ignore it. In the following, we  
call this quantity the intrinsic dissipation of the driving process. 
Then $\varphi = \varphi(F_{\I},F_{\II})$ is the 
relative intrinsic dissipation of the output process with respect to the driving process, and
finally $\xi=\xi(F_{\I},F_{\II})$ quantifies the \emph{degree of coupling}  
\cite{Kedem1965_vol61, Polettini2015_vol114, Caplan1966_vol10, Esposito2009_vol102, Benenti2017_vol694}. 
From the constraints of Eqs.~(\ref{eq:def-positive}-\ref{eq:contrainteEngine}), these functions are bounded by 
\begin{equation}
\Pi \geq 0, \quad \xi \in [-1,0[, \quad \varphi \in [0,-\xi],
\end{equation}
for the system to operate as a machine. If it does not, the above parametrization could still be used but with 
a modified range of the parameters, namely $\varphi\geqslant 0 $ and $\xi \in [-1,1]$. 
Note that we have also excluded the value $\xi=0$ from our analysis which corresponds to having independent driving and output processes for which $G_{\I,\II}=0$. In this case, the system cannot work as a machine because its 
efficiency would be negative with $\eta = - \varphi^2 \leq 0$. 
Note also that in the literature on thermoelectricity 
\cite{Kedem1965_vol61, Benenti2017_vol694}, 
it is customary to use the figure of merit
$ZT$ instead of the degree of coupling. The two notions are simply related by 
$ZT=\xi^2/(1- \xi^2)$, so that $ZT$ is a real positive number which goes to infinity when $\xi$ 
tends to $-1$.

Restricting ourselves to a working machine, 
we use Eqs.~(\ref{eq:NonEqRespMatrix}--\ref{eq:NonEqRespMatrix2}) in the definition~(\ref{eq:efficiency}) of thermodynamic efficiency to obtain 
\begin{equation}
  \label{eq:effGij}
  \eta = -\f{G_{\I,\II}F_{\I}F_{\II}+ G_{\II,\II}F_{\II}^{2}}{G_{\I,\I} F_{\I}^2 +G_{\I,\II}F_{\I}F_{\II}},
\end{equation}
which can be turned into 
\begin{equation}
  \label{eq:defEff}
\eta = -\f{\varphi^2+\xi\varphi}{1+\xi\varphi},
\end{equation}
with the aid of Eq.~(\ref{eq:new-parameters}). We emphasize that with this new parametrization, the machine efficiency 
does not to depend explicitly on the intrinsic dissipation $\Pi$, but only depends on the relative intrinsic dissipation $\varphi$ and on 
the degree of coupling $\xi$. The specific dependence 
of the efficiency on the affinities is then completely transferred to $\varphi$ and $\xi$. 
As we shall see below, this new parametrization of the efficiency provides useful insights into the machine properties. One important benefit in particular is the ability to bound the machine efficiency and output power.

\subsection{Tight coupling far from equilibrium}
\label{sec:tightcoupling}

In this section, we discuss the notion of tight coupling  far-from-equilibrium based on the non-equilibrium conductance matrix and the $(\Pi,\varphi,\xi)$ parametrization. Tight coupling between two entropy fluxes means that the elementary steps must produce entropy in constant proportion. In other words, the physical quantities corresponding to the driving and output processes must be always exchanged in the same proportion in such a way that the two equations in~(\ref{eq:NonLinearRegime}) are linearly dependent. 
The latter condition implies that the matrix $\bm{G}$ is of rank one, which means that it can be written in the form
\begin{equation}
	\bm{G} = 
	\left( \begin{array}{cc} 
	G_{\I, \I} & G_{\I, \II} \\
	G_{\II, \I} & G_{\II, \II} \\
	\end{array} \right) = G_{\I, \I} \left( \begin{array}{cc} 
	1 & \alpha \\
	\alpha & \alpha^{2} \\
	\end{array} \right),
\end{equation}
in terms of a real coefficient $\alpha$.
Using Eq. (\ref{eq:NonLinearRegime}), one finds $\Jmoyi{\I}= \alpha \Jmoyi{\II}$, thus $\alpha$ is precisely the proportionality factor between the two currents. Then using Eq. (\ref{eq:GTC}), one finds $\bm{G}=\bm{G}_{min}$. Thus $\bm{G}_{min}$ is the non-equilibrium conductance matrix of the system if it operates in the tight coupling regime. Furthermore, this shows that the inequality of Eq. (\ref{Gmin}) becomes saturated in the tight coupling regime.

Now, from Eqs.~(\ref{eq:new-parameters}) and (\ref{eq:defEff}), the coupling parameter 
reaches the value $\xi = \text{sign}(F_{\I}F_{\II}\alpha) = -1$, 
because $\xi \in [-1,0[$, and $\eta = \varphi =  | \alpha F_{\II}/F_{\I} |$. Thus, in 
the tight coupling regime, the degree of coupling reaches its minimum value.

Going back to the general case, one deduces from Eq.~(\ref{eq:defEff}) that
\begin{equation}
\left. \frac{\partial \eta}{\partial \xi} \right|_{\varphi}=-\frac{\varphi (1- \varphi^2)}{(1+ \xi \varphi)^2},
\end{equation}
which is always negative since $\varphi \in [0,1]$.
Therefore, the efficiency monotonously increases when $\xi$ decreases, and the maximum value of the efficiency at 
fixed value of $\varphi$ is reached when $\xi=-1$, i.e. at tight coupling.

\subsection{Maximum efficiency as function of the degree of coupling}

We now bound the efficiency $\eta=\eta(\xi,\varphi)$ of Eq.~(\ref{eq:defEff}) by looking  
at the value of the function $\varphi$ that yields the maximum efficiency in Eq~(\ref{eq:defEff}) at a fixed degree of coupling $\xi$.
The condition $\left. \partial \eta / \partial \varphi \right |_{\xi} =0$ leads to a simple second degree polynomial equation 
\begin{equation} \label{eq:polynomial}
	\xi \varphi^{2} + 2\varphi + \xi =0 .
\end{equation}
Multiplying  the numerator and denominator of Eq.~(\ref{eq:defEff}) by $2+\xi\varphi$ and 
using (\ref{eq:polynomial}), we find that the maximum efficiency becomes $\eta_\mathrm{max} = - \xi \varphi /(2+\xi\varphi)$. 
Using the solution of Eq.~(\ref{eq:polynomial}) in this expression of $\eta_\mathrm{max}$, 
we obtain the maximal machine efficiency in terms of the degree of coupling function $\xi$, 
\begin{equation}
\label{eq:Max-Eff}
\eta_{\text{max}}(\xi)\equiv \f{1-\sqrt{1-\xi^2}}{1+\sqrt{1-\xi^2}},
\end{equation}
which is such that 
\begin{equation}
\eta_{\text{max}}(\xi) \geq \eta(\xi,\varphi)
\label{eq2:Max-Eff}
\end{equation} for all $\xi$ and $\varphi$ in the allowed range.
This inequality is illustrated in Fig.~\ref{fig:etaxi} for the model of molecular motor studied in section \ref{sec:Gij-models}.
As expected from the previous section, Eq.~(\ref{eq:Max-Eff}) also confirms 
that the maximum of the curve $\eta_{\text{max}}(\xi)$ is reached 
when the condition of tight coupling holds namely $\xi=-1$ since at this point $\eta_{\text{max}}=1$. 

Since the maximum efficiency depends only on the degree of coupling $\xi$, it is possible to bound the 
efficiency by measuring the degree of coupling. For instance, if it is known that $\xi_\text{min} \leqslant \xi$ for all conditions of operation of the machine, then we can deduce from Eq.~(\ref{eq2:Max-Eff}) that
  $\eta \leqslant \eta_{\text{max}}(\xi_\text{min})$.
Note that the bound itself is not unique because 
$\xi$ is constructed from the non-equilibrium conductance matrix 
which is not uniquely defined by Eq.~(\ref{eq:NonLinearRegime}); 
nevertheless the dependance of $\eta_{\text{max}}$ versus $\xi$ is universal.

\subsection{Power-efficiency relations}
\label{sec:power-eff}

In this section we derive two upper bounds for the entropy production rate of the output process, a quantity which is the product of the output power of the machine with its affinity. These bounds are functions of the efficiency and hence are called power-efficiency relations, since they represent a constraint for reaching both high power and high efficiency.

To obtain the first bound, we factorize $G_{\I,\I}F_{\I}^{2}$ in Eq.~(\ref{eq:NonEqRespMatrix2}):
\begin{equation}
	- \sigma_{\II} = - G_{\I,\I}F_{\I}^{2} \left( \frac{G_{\I,\II}F_{\II}}{G_{\I,\I}F_{\I}} + \frac{G_{\II,\II}F_{\II}^{2}}{G_{\I,\I}F_{\I}^{2}} \right) = - \Pi \left( \xi \varphi + \varphi^{2} \right).
\end{equation}
From Eq.~(\ref{eq:defEff}) we have $-(\xi \varphi + \varphi^{2}) = \eta (1+\xi \varphi)$ and therefore
\begin{equation}
- \sigma_{\II} = \Pi \eta \left( 1 + \xi \varphi \right),
\label{s2}
\end{equation}
then using again Eq.~(\ref{eq:defEff}), we can express $\varphi$ in terms of $\eta$ and $\xi$ as
\begin{equation}
\varphi^\pm = - \frac{\xi \left( \eta + 1 \right)}{2} \pm \frac{1}{2} \sqrt{(\eta + 1)^2 \xi^2 - 4 \eta},  
\end{equation}  
where we have used Eq.~(\ref{eq:Max-Eff}-\ref{eq2:Max-Eff}) to guarantee that $\varphi$ is real.  
Inserting these two solutions in Eq.~(\ref{s2}), we obtain
\begin{equation}
-\sigma_{\II}^\pm = \Pi \eta\left(1-\xi^2\f{1+\eta}{2} \mp \xi \sqrt{\f{\xi^2}{4}(1+\eta)^2-\eta} \right).
\end{equation} 
This equation shows that the relation between the output entropy production rate $-\sigma_2$ and the efficiency is in general 
bi-valued, which means that there are two possible values of the output entropy production rate for the same value of the efficiency. 
This relation becomes single-valued when $\xi=-1$, i.e. for tight coupling, since in this case $-\sigma_{2}^{-}$ is equal to zero, and only  $-\sigma_2^+$ remains.

In the general case of arbitrary coupling, it is enough to upper bound $-\sigma_2^+$ to obtain a general bound on the output entropy production rate because $-\sigma_2^+ \geq -\sigma_2^-$ for $\xi \in [-1,0[$. Since one can also show that $-\sigma_2^+$ is always a decreasing function of $\xi$ at fixed $\eta$, its maximum value is reached at $\xi=-1$, which corresponds to the tight coupling condition. When inserting $\xi=-1$ into the expression of $-\sigma_2^+$, we obtain the first inequality:
\begin{equation}
  \label{eq:PowerEffS1}
-\sigma_{\II} \leq  \Pi \eta (1-\eta).
\end{equation}

Alternatively, one can start from Eq.~(\ref{eq:NonEqRespMatrix2}) and factorize $G_{\II,\II} F_{\II}^2$ which leads to 
$\sigma_{\II} = G_{\II,\II} F_{\II}^2(1+\xi/\varphi)$. Then, using again the explicit solution of $\varphi$ 
as a function of $\xi$ and $\eta$, one obtains an expression 
which when evaluated at $\xi = -1$ leads to the second inequality
\begin{equation}   \label{eq:PowerEffS2}
	-\sigma_{\II}  \leq G_{\II,\II} F_{\II}^2 \f{1-\eta}{\eta}.
\end{equation}
Despite the apparent similarities between Eqs.~(\ref{eq:PowerEffS1}) and (\ref{eq:PowerEffS2}) 
with the bounds recently derived in Ref.~\cite{Pietzonka2017_vol}, 
we would like to stress that Eqs.~(\ref{eq:PowerEffS1}) and (\ref{eq:PowerEffS2}) represent a different result because 
our bounds are based on the non-equilibrium conductance matrix using classical thermodynamics in a macroscopic and deterministic setting.
In contrast to that, the bounds of Ref.~\cite{Pietzonka2017_vol} have been derived in a stochastic setting
based on uncertainty relations. The introduction of the non-equilibrium conductance matrix, 
the parametrization of the efficiency and Eqs.~(\ref{eq:PowerEffS1}--\ref{eq:PowerEffS2}) represent our first main results. 
In the following, we explain how to reconcile both bounds 
within a formalism of large deviation of currents.

\section{Construction of the non-equilibrium conductance matrix from a large deviation function framework}
\label{sec:quadr-bounds-large}
So far, our analysis was based primarily on classical thermodynamics, where currents are deterministic quantities.
In contrast to that, we introduce in the following a stochastic thermodynamics description, in which currents
become random variables. 
As far as the microscopic dynamics is concerned, we assume that it 
can be described as a Markov jump process. This Markovian dynamics admits a non-equilibrium stationary state. 
In the following, by exploiting a 
 quadratic bound of the LDF of currents near this non-equilibrium stationary state, we 
 show how to define uniquely the NE conductance matrix.

 We denote with upper case letters average quantities 
and with lower case letters the corresponding fluctuating quantities; then the subscript indicates the level of description, 
$(x,y)$ for an edge, $c=c_1,c_2,\dots$ for cycles and $\X=1,2$ for the physical quantities.

\subsection{Physical, cycle and edge currents and affinities}
\label{sec:micr-model-mach}

The probability per unit time to jump from the state $y$ to state $x$ of the machine is given by the rate matrix 
of components $k_{\exy} \geq 0$. We call the couple of states $\exy$ an oriented edge when $k_{\exy} > 0$. 
We assume that if the jump from $y$ to $x$ is possible then the reverse jump also exists, i.e. $k_{\exy} > 0$ implies that $k_{\eyx} > 0$.  The stationary probability of $x$, 
denoted $\pi_{x}$, verifies by definition $\sum_{y} k_\exy \pi_{y} = 0$.  
The number of oriented edges is $|E|$. The average probability current along edge $\exy$ in the stationary state is
\begin{equation}
  \label{eq:EdgeCurrent}
	\Jmoyi{\exy} \equiv k_{\exy} \pi_{y} - k_{\eyx} \pi_{x},
\end{equation}
 and the edge affinity 
\begin{equation}
  \label{eq:Edgedef}
    F_{\exy} \equiv \ln \f{k_{\exy} \pi_{y}}{ k_{\eyx} \pi_{x}}.
  \end{equation}

  Another level of description is that of cycles 
which consist of several edges connected together. Cycle  
currents are linearly connected to edge currents \cite{Book_Hill1989} as:
  \begin{equation}
    \label{eq:cyclecurrent2edge}
    \Jmoyi{\exy} \equiv \sum_{c \in C} A_{\exy,c}J_c,
  \end{equation}
  where $\bm{A}$ is an $|E| \times |C|$ matrix, such that 
$A_{\exy,c}$ is zero if the edge $(x,y)$ does not belongs to the cycle $c$, 
or $\pm1$ if it belongs to it with the sign providing the orientation. 
The columns of $\bm{A}$ form a basis of vectors called fundamental cycles, 
and the ensemble of fundamental cycles is denoted $C$ with cardinal $|C|$.
This set is called fundamental because it is a minimal set 
of linearly independent cycles that can generate any cycle of 
the graph \cite{Schnakenberg1976_vol48}.

Each physical thermodynamic force involved in 
the interaction of the machine with its environment has 
a corresponding physical (also called sometimes operational for this reason) current
associated with it. These currents are also linearly related to the cycle 
currents as:
\begin{equation}
  \label{eq:fundamentalcylecurrent}
  \Jmoyi{\X} \equiv \sum_{c\in C} \phi_{\X,c} \Jmoyi{c}, 
\end{equation}
where $\phi_{\X,c}$ represents the amount of the  
physical quantity $\X$ which is exchanged with the environment 
when the cycle $c$ is run once. Note that by construction the coefficients of the 
matrix $\bm{\phi}$ are dimension full, depending on the choice of physical currents, 
unlike the coefficients of the matrix $\bm{A}$ which are dimensionless.
  
The entropy production takes the same value 
on these three levels, thus
\begin{equation}
    \label{eq:sigmaedgecycle}
\sigma = \sum_{\exy}\Jmoyi{\exy} F_{\exy}   =\sum_{c}J_c F_c = \sum_{\X} J_{\X}F_{\X}.
\end{equation}
This allows to connect the edge, cycle and physical affinities through dual forms 
of Eqs.~(\ref{eq:cyclecurrent2edge}--\ref{eq:fundamentalcylecurrent})\cite{Book_Hill1989,Polettini2016_vol94}:
\begin{eqnarray}
	F_{c} &=& \sum_{\exy} F_{\exy} A_{\exy,c}, \label{eq:edge2cycleAff} \\
	F_{c} &=& \sum_{\X} F_{\X} \phi_{\X,c} \label{eq:cycle2physAff}.
\end{eqnarray}

\subsection{Quadratic bound on large deviations}
\label{sec:bounds-large}

In a stochastic description of the machine, all the currents introduced above 
at the various levels become stochastic quantities.
Let us denote $\Jstochi{\exy}$ as the edge current associated to  
the net number of transitions from $y$ to $x$ per unit time during 
a trajectory of duration $t$. 
These edge currents $\{ \Jstochi{\exy} \}$ are fluctuating quantities which 
are assumed to obey a large deviation principle. This means that the probability $P$ of observing them in a total time $t$ decays as 
\begin{equation}
  \label{eq:LDFprob}
  P(\{ \Jstochi{\exy} \}) \simeq_{t \to +\infty} e^{-t I( \{ \Jstochi{\exy} \} )  },
\end{equation}
where $I( \{ \Jstochi{\exy} \} )$ is the large deviation function (LDF) of the currents also called rate function \cite{Touchette2009_vol478}.

After some manipulations of Eq.~(3) of Ref.~\cite{Gingrich2016_vol116} 
\footnote{More precisely, one needs to express the term $\sigma^\pi(y,z)$ of Eq.~(3) of Ref.~\cite{Gingrich2016_vol116} as $j^\pi(y,z)F(y,z)$ and divide $j^\pi(y,z)$ out to obtain Eq.~(\ref{eq:Dissipationboundscurrent}).}, a quadratic bound 
for the LDF of edge currents can be written in the form
\begin{equation}
  \label{eq:Dissipationboundscurrent}
  I(\{\Jstochi{\exy}\}) \leq \f{1}{4} \sum_{\exy} (\Jstochi{\exy} -  \Jmoyi{\exy} )^2 \bar R_\exy,
\end{equation}
where 
\begin{equation} \label{eq:edgeResistance}
	\bar R_\exy \equiv \frac{F_\exy}{  \Jmoyi{\exy}}
\end{equation}
represents the components of a diagonal resistance matrix, i.e. an edgewise resistance. 
Now, the cycle currents $\Jstochi{c}$ are connected to the edge currents $\Jstochi{\exy}$ by the stochastic version of Eq.~(\ref{eq:cyclecurrent2edge}).
Let us introduce $\bm{\tilde j} \equiv (j_{c_{1}},j_{c_{2}},\dots,j_{c_{|C|}})^{T}$ the vector of the cycle currents and $\bm{\tilde J}$ its mean value. When using Eq.~(\ref{eq:cyclecurrent2edge}) as a change of variable into Eq.~(\ref{eq:Dissipationboundscurrent}), we obtain 
\begin{equation}
  \label{eq:boundsLDFcycle}
  I(\bm{\tilde j})\leq \f{1}{4} (\bm{\tilde j} -  \bm{\tilde J})^{T} \cdot \bm{\tilde R} \cdot (\bm{\tilde j} -  \bm{\tilde J}) ,
\end{equation}
where $ \bm{\tilde R} $ is the cycle
resistance matrix  of components 
\begin{equation} \label{eq:CycleResistanceMatrix}
    \tilde R_{c,c'} \equiv \sum_\exy  A_{\exy,c}A_{\exy,c'} \f{F_\exy}{ \Jmoyi{\exy}}.
\end{equation} 

By contracting Eq.~(\ref{eq:boundsLDFcycle}) over cycle currents,  
one obtains an upper bound for the LDF of physical currents $\Jstochi{\I},\Jstochi{\II}$. The LDF we are interested in reads
\begin{equation}
  \label{eq:miniCycleLDF}
  I_\text{quad} ( \Jstoch ) = \frac{1}{4} \min_{ \{..\} } \left( \bm{\tilde j} -  \bm{\tilde J} \right)^{\rm T} \cdot 
\bm{\tilde R} \cdot \left(\bm{\tilde j} -  \bm{\tilde J}\right),
\end{equation}
where $\{..\}$ denotes the minimum over currents $\bm{\tilde j}$ such that $\Jstoch=\bm{\phi} \cdot\bm{\tilde j}$, with $\Jstoch$ the vector of 
physical currents $(\Jstochi{1},\Jstochi{2})$.
Since the function to be minimized is quadratic and the constraints 
are linear, this contraction can be achieved exactly as follows:
The function to be minimized is 
\begin{equation}
f_\text{quad}= \frac{1}{4} \left( \bm{\tilde j} - \bm{\tilde J} \right)^{\rm T} \cdot 
\bm{\tilde R} \cdot \left( \bm{\tilde j} - \bm{\tilde J} \right) - \bm{\lambda}^{\rm T} \cdot \left(\Jstoch-\bm{\phi} \cdot \bm{\tilde j}\right),
\end{equation} 
where $\bm{\lambda}$ is a Lagrange multiplier. After minimizing $f_\text{quad}$ with respect to $\bm{\tilde j}$, 
one obtains an expression of $\bm{\tilde j}$ as a function of $\bm{\lambda}$. Then using 
again the constraint $\Jstoch=\bm{\phi} \cdot\bm{\tilde j}$, one finds
\begin{equation}
\bm{\lambda}= - \frac{1}{2} \left[ \bm{\phi} \cdot \bm{\tilde R}^{-1} \cdot \bm{\phi}^{\rm T} \right]^{-1} \cdot 
\left( \Jstoch - \Jmoy \right).
\end{equation}
Inserting this expression into $\bm{\tilde j}$ and using it into $I_\text{quad}$, one obtains
\begin{equation}
\label{eq:Iquad}
I_\text{quad} ({\bm j}) =   \frac{1}{4} \left( \bm{j} - \bm{J} \right)^{\rm T} \cdot 
\bm{R} \cdot \left( \bm{j} - \bm{J} \right).
\label{I_quad}
\end{equation}
where we have introduced $\bm{R}$ as the $2 \times 2$ resistance matrix in the basis of physical currents
\begin{equation}
  \label{eq:defGLDF}
  \bm{R}  \equiv \left( \bm{\phi} \cdot  \bm{\tilde R}^{-1} \cdot \bm{\phi}^{\rm T} \right)^{-1}.
\end{equation}
In the end, we obtain the following inequality for the LDF of physical currents:
\begin{equation}
  \label{eq:quad-boundsG}
I (\Jstoch ) \leq I_\text{quad} (\Jstoch).
\end{equation}

The quadratic bound on the LDF used in Eq. (\ref{eq:Dissipationboundscurrent}) has been  
built to respect the fluctuation theorem \cite{Seifert2012_vol75, Gingrich2016_vol116}. Therefore, at the level of physical observables the quadratic bound obeys the relation
\begin{equation}
  \label{eq:fluctuationThmquad}
   I_\text{quad} ({\Jstoch})-I_\text{quad} ({-\Jstoch})= - \Jstoch^{\rm{T}} \cdot \bm{F}.
 \end{equation}
Once Eq.~(\ref{eq:Iquad}) is inserted into this equation, we obtain
$\Jstoch^{\rm{T}}  \cdot \bm{R} \cdot \Jmoy= \Jstoch^{\rm{T}} \cdot \bm{F}$ for all $\bm{j}$, or equivalently
 $\bm{R} \cdot \Jmoy=\bm{F}$. After comparing with Eq. (\ref{eq:NonLinearRegime}),
we deduce the relation
 \begin{equation}
   \label{eq:GegalRmoinsun}
    \bm{G} = \bm{R}^{-1} = \bm{\phi} \cdot  \bm{\tilde R}^{-1} \cdot \bm{\phi}^{\rm T},
 \end{equation}
that provides a consistent definition of the conductance matrix. Also note that we have assumed the matrices $\bm{\tilde R}$, $\bm{R}$ and $\bm{G}$ to be invertible, if it is not the case the matrix $\bm{G}_{min}$ introduced in Eq.~(\ref{eq:GTC}) should be used from the beginning. 
An explicit example of this case is provided for an unicyclic machine in the section~\ref{sec:unicyclic-engine-ue}.

In appendix \ref{MicroToMacro}, we derive an alternate route leading to Eq.~(\ref{eq:GegalRmoinsun}), which avoids the last step of Eq.~(\ref{eq:fluctuationThmquad}) but relies instead on a further change of the level of description from that of cycles to that of physical currents.

To summarize, the property that edge current fluctuations in non-equilibrium stationary states are 
more likely than those predicted by 
linear response analysis \cite{Pietzonka2016_vol93, Gingrich2016_vol116} which is Eq.~(\ref{eq:Dissipationboundscurrent}), 
carries out to the level of cycles and from there to the level of physical macroscopic currents. This approach leads to a relation 
between affinities and physical macroscopic currents that defines the non-equilibrium conductance matrix. The construction of this matrix
from a large deviation framework represents our second main result.

\section{Implications for the thermodynamic efficiency and the output power}
\label{sec:boundsEff}
In this section, we show how previously obtained bounds on efficiency \cite{Pietzonka2016_vol2016}, and power-efficiency
 trade-offs \cite{Pietzonka2017_vol} can be derived from a framework based on the non-equilibrium conductance matrix.

\subsection{Inequality involving the non-equilibrium conductance matrix and the covariance matrix of physical currents}
\label{sec:boundsUncertaintyRelations}

We introduce the cumulant generating function (CGF) defined by
\begin{equation}
	\lambda(\gamma_{\I},\gamma_{\II}) \equiv \lim_{t \rightarrow \infty}\frac{1}{t} \ln \left \langle e^{t \left ( \gamma_{\I} \Jstochi{\I} + \gamma_{\II} \Jstochi{\II} \right) } \right \rangle,
\end{equation}
which is the Legendre transform of the LDF of the physical currents 
$\lambda(\gamma_1,\gamma_2) = \max_{j_\I,j_\II} [(\gamma_1 j_1+\gamma_2 j_2-I(j_1,j_2) ]$. 
Similarly, the Legendre transform of the quadratic LDF is
$\lambda_{\text{quad}}(\gamma_1,\gamma_2) \equiv \max_{j_\I,j_\II} [(\gamma_1 j_1+\gamma_2 j_2-I_\text{quad}(j_1,j_2) ]$. From Eq.~(\ref{eq:quad-boundsG}), we have
\begin{equation}
  \label{eq:CGFquad}
  \lambda_{\text{quad}}(\gamma_1,\gamma_2) \leq \lambda(\gamma_\I,\gamma_\II),
\end{equation}
where $\lambda_{\text{quad}}(\gamma_1,\gamma_2)$ can be explicitly determined using Eq.~(\ref{I_quad}). The maximum 
with respect to $\Jstochi{1}$ and $\Jstochi{2}$ leads to the condition 
\begin{equation}
  \label{eq:gammminCGF}
\bm{\gamma} = \frac{1}{2} \bm{R} \cdot \left( \Jstoch - \Jmoy \right),
\end{equation}
where $\bm{\gamma}$ is the vector $(\gamma_1,\gamma_2)$.
Inserting this result into the definition of $\lambda_{\text{quad}}$ and using the property 
$\bm{R}^{-1} = \bm{G}$, we obtain
\begin{equation}
  \label{eq:boundsCGF}
    \lambda_{\text{quad}}(\bm{\gamma}) =\bm{\gamma}^{\rm T} 
    \cdot \bm{G} \cdot \bm{\gamma} +  \Jmoy\cdot \bm{\gamma}.
\end{equation}
This equation holds for any value of the conjugated variables $\bm{\gamma}$.
Then, the functions $\lambda$ and $\lambda_{\text{quad}}$ have the same  value at origin and the same first derivative with respect to $\bm{\gamma}$ around the origin, 
therefore the inequality (\ref{eq:CGFquad}) can be carried out to second order derivatives.
The result is the following inequality
\begin{equation}
  \label{eq:ineqHessian}
\forall \bm{\gamma} \in \mathbb{R}^2, \quad\bm{\gamma}^{\rm T} 
    \cdot \bm{G} \cdot \bm{\gamma} \leq \frac{1}{2} \bm{\gamma}^{\rm T} 
    \cdot \bm{C} \cdot \bm{\gamma},
\end{equation}
where we have introduced the covariance matrix as
\begin{equation}
  C_{\X\Y}=\text{Cov}(\Jstochi{\X},\Jstochi{\Y}) \equiv \lim_{t \to \infty} t\left[  \left \langle \Jstochi{\X} \Jstochi{\Y}  \right \rangle - \left \langle \Jstochi{\X} \right \rangle \left \langle \Jstochi{\Y}  \right \rangle \right]= \frac{\partial^{2}\lambda}{\partial \gamma_{\X} \partial \gamma_{\Y}}(0,0).
\end{equation}

Now, Eq. (\ref{eq:ineqHessian}) simply reads $\bm{G} \leq \bm{C}/2$ for the matrix order introduced in Eq.~(\ref{eq:loewnerpartialorder}).
Choosing $\bm{\gamma} = (\gamma_{\I},0)^{\rm T}$ or $(0,\gamma_{\II})^{\rm T}$ in Eq.~(\ref{eq:ineqHessian}) leads to the tight bounds derived in \cite{Polettini2016_vol94a}:
\begin{eqnarray}
  \label{eq:tightbounds1}
  G_{\I,\I} {F_{\I}}^2 &\leq& \f{\mbox{Var}(\sigma_{\I})}{2}, \\
  G_{\II,\II} {F_{\II}}^2 &\leq& \f{\mbox{Var}(\sigma_{\II})}{2},   \label{eq:tightbounds2}
\end{eqnarray}
after multiplying the inequalities by ${F_{\I}}^{2}$ or ${F_{\II}}^{2}$ respectively.
These inequalities are saturated in the linear regime close to equilibrium, where the non-equilibrium conductance matrix becomes the 
standard Onsager matrix $\bm{L}$ and the relation $\bm{L} = \bm{C}/2$ is the well-known fluctuation-response relation.
When using the above inequalities (\ref{eq:tightbounds1}--\ref{eq:tightbounds2}) into  (\ref{eq:PowerEffS1}--\ref{eq:PowerEffS2}), one obtains
\begin{eqnarray}
\label{eq:combine}
- \langle \sigma_{\II} \rangle &\leq& \Pi \eta(1-\eta) \leq \f{\mbox{Var}(\sigma_{\I})}{2} \eta(1-\eta), \\
- \langle \sigma_{\II} \rangle &\leq& G_{22} (F_2)^2 \f{1-\eta}{\eta} \leq \f{\mbox{Var}(\sigma_{\II})}{2} \f{1-\eta}{\eta}.
\end{eqnarray}
Thus we retrieve the power-efficiency trade-offs derived by Pietzonka and Seifert~\cite{Pietzonka2017_vol}
\begin{eqnarray}
  \label{eq:PowerEffTradeOffVariance}
- \langle \sigma_{\II} \rangle &\leq& \f{\mbox{Var}(\sigma_{\I})}{2} \eta(1-\eta),\\
  \label{eq:PowerEffTradeOffVariance2}
- \langle \sigma_{\II} \rangle &\leq& \f{\mbox{Var}(\sigma_{\II})}{2} \f{1-\eta}{\eta}.
\end{eqnarray}
\subsection{From uncertainty relations to bounds on the efficiency}
\label{sec:tight-coupling-limit}

By combining the inequality $\bm{G} \leq \bm{C}/2$  obtained in the previous section 
with Eq. (\ref{Gmin}), one obtains
\begin{equation}
\bm{G}_{min} \leq \bm{G} \leq \f{\bm{C}}{2},
\end{equation}
where the first inequality on the left hand side becomes 
saturated only if the system has strongly coupled physical currents. Using again the property (\ref{eq:loewnerpartialorder}) for the matrix order, 
 the relation $\bm{G}_{min} \leq \bm{C}/2$ implies three  
inequalities by choosing three particular values of the vector $\bm{x}$, namely $(F_\I,0)^{\rm T}$, $(0,F_\II)^{\rm T}$ and $(F_\I,F_\II)^{\rm T}$.
These are the so-called \emph{uncertainty relations} \cite{Pietzonka2016_vol93,Gingrich2016_vol116}:
\begin{eqnarray}
  \label{eq:uncertaintyrelations}
  \f{ \langle \sigma_{\I} \rangle ^2}{ \langle \sigma \rangle } &\leq& \f{\mbox{Var}(\sigma_{\I})}{2}, \\
  \label{eq:uncertaintyrelations2}
  \f{ \langle \sigma_{\II} \rangle ^2}{ \langle \sigma \rangle} &\leq& \f{\mbox{Var}(\sigma_{\II})}{2},\\
  \label{eq:uncertaintyrelations3}
  \langle \sigma \rangle & \leq&  \f{\mbox{Var}(\sigma)}{2}.
\end{eqnarray}

Now, we recall the definition of efficiency in terms of average partial and total entropy production rates 
\begin{equation}
    \label{eq:relationsentropieEff}
     \eta =  1-\frac{\langle \sigma \rangle}{\langle \sigma_\I \rangle}.
\end{equation}
Inserting this definition into Eqs.~(\ref{eq:uncertaintyrelations})-(\ref{eq:uncertaintyrelations3}), 
one recovers two known bounds on efficiency ~\cite{Pietzonka2016_vol2016}:
\begin{equation}
\label{eq:upperboundsEff}
\eta \leqslant \min \left(1- 2 \f{\langle \sigma_{\I} \rangle}{\mbox{Var}(\sigma_{\I})}, 
\f{1}{1 -2 \f{\langle \sigma_{\II} \rangle}{\mbox{Var}(\sigma_{\II})}} \right),
\end{equation}
and
\begin{equation}
\label{eq:lowerboundsEff}
\eta \geqslant \max  \left( 1-\f{\mbox{Var}(\sigma)}{\langle \sigma_{\I} \rangle} , 
\f{1}{1-\f{\mbox{Var}(\sigma)}{\langle \sigma_{\II} \rangle} }\right).
\end{equation}
Among these two inequalities, the first one namely Eq.~(\ref{eq:upperboundsEff}) is 
probably the most useful one because it involves only the partial entropy production
rates of process $1$ or $2$, whereas Eq.~(\ref{eq:lowerboundsEff}) requires 
information on both processes which is often missing.

\section{Illustration on small machines}
\label{sec:Gij-models}

\subsection{Unicyclic Engine}
\label{sec:unicyclic-engine-ue}

\begin{figure}[ht]
  \centering  \includegraphics[scale=0.6]{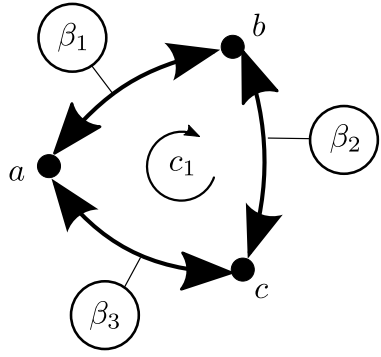}
  \caption{Sketch of the unicyclic heat to heat converter with three states $a$, $b$, and $c$. Transition $a \leftrightarrow b$ is promoted by the heat reservoir at inverse temperature $\beta_{1}$, transition $b \leftrightarrow c$ by the heat reservoir at inverse temperature $\beta_{2}$, and $c \leftrightarrow a$ for $\beta_{3}$.}
  \label{fig:sketchunicyclic}
\end{figure}

We start by studying a simple example of heat-to-heat converter. We consider the unicyclic three states 
model depicted on figure~\ref{fig:sketchunicyclic}. Each state $a,b,c$ has 
a different energy $E_a,E_b,E_c$ and each transition is connected to a different heat reservoir at inverse temperature $\beta_1,\beta_2,\beta_3$. The transition rates are
\begin{equation}
  \label{eq:ratesUE}
  \begin{array}{ll}
    k_{(b,a)}= \Gamma e^{-\f{\beta_1}{2}(E_b-E_a)},&k_{(a,b)}= \Gamma e^{-\f{\beta_1}{2}(E_a-E_b)},\\
    k_{(c,b)}= \Gamma e^{-\f{\beta_2}{2}(E_c-E_b)},&k_{(b,c)}= \Gamma e^{-\f{\beta_2}{2}(E_b-E_c)},\\
    k_{(a,c)}= \Gamma e^{-\f{\beta_3}{2}(E_a-E_c)},&k_{(c,a)}= \Gamma e^{-\f{\beta_3}{2}(E_c-E_a)},\\
  \end{array}
\end{equation}
where $\Gamma$ is the coupling strength to the reservoirs. The system is coupled to three heat reservoirs, and 
its total entropy production rate is $\sigma = -\beta_1 \Jmoyi{1} - \beta_2 \Jmoyi{2} - \beta_3 \Jmoyi{3}$, where $\Jmoyi{i}$ denotes the heat flux from the heat reservoir $i$ to the system. Using the energy conservation $\Jmoyi{1}+\Jmoyi{2}+\Jmoyi{3}=0$, we obtain the total entropy production rate $\sigma =  (\beta_3-\beta_1)\Jmoyi{1} + (\beta_3-\beta_2)\Jmoyi{2}$. We consider as driving current the heat flow $\Jmoyi{1}$ and output current the heat flow $\Jmoyi{2}$, and we assume that the temperatures of the reservoirs satisfy $\beta_3 > \beta_1$ and $\beta_3>\beta_2$ and the energies are such that $E_b> E_c > E_a$. Under these conditions, the driving and output currents are such that $\Jmoyi{1}>0$ and $\Jmoyi{2}<0$, 
and the system operates as a machine that transfers heat 
from a cold to a hot reservoir using the thermodynamic force generated by the 
transfer of heat from a hot to a cold reservoir.
The partial entropy production rates and physical affinities are then
\begin{equation}
  \label{eq:PhysicalUE}
  \begin{array}{ll}
    \sigma_1 = (\beta_3-\beta_1)\Jmoyi{1},& F_1=(\beta_3-\beta_1)\\
     \sigma_2 = (\beta_3-\beta_1)\Jmoyi{2},& F_2=(\beta_3-\beta_2).
  \end{array}
\end{equation}

At the lower level, the system has a single cycle $c_{1}$ for which we chose the orientation $a\to b\to c$. Thus, the matrix of fundamental cycles $\bm{A}$ is actually the vector $\bm{A}=(1,1,1)^{\rm T}$. Due to the stationary condition, the current is the same on each edge and is equal to the cycle current
\begin{equation}
  \label{eq:cyclecurrentUE}
  \Jmoyi{(b,a)}=\Jmoyi{(c,b)}=\Jmoyi{(a,c)}= \Jmoyi{c_1} = \f{\Gamma}{Z} \left(k_{(b,a)}k_{(a,c)}k_{(c,b)} -k_{(a,b)}k_{(b,c)}k_{(c,a)}\right),
\end{equation}
where we have defined
\begin{eqnarray}
Z&=&k_{(a,b)}k_{(a,c)}+k_{(a,b)}k_{(b,c)}+k_{(a,c)}k_{(c,b)}+k_{(b,a)}k_{(b,c)}+k_{(b,c)}k_{(c,a)} \nonumber \\ &&+k_{(b,a)}k_{(a,c)}+k_{(c,a)}k_{(c,b)}+k_{(c,b)}k_{(b,a)}+k_{(c,a)}k_{(a,b)}.
\end{eqnarray}
The corresponding edge affinities are defined in Eq.~(\ref{eq:Edgedef}). From the matrix of fundamental cycles $\bm{A}$, we derive the cycle affinity
\begin{equation}
  \label{eq:cycleaffinityUE}
  F_{c_1}= (\beta_3-\beta_1) (E_b-E_a) + (\beta_3-\beta_2)(E_c-E_b)= \ln \f{k_{(b,a)}k_{(a,c)}k_{(c,b)}}{k_{(a,b)}k_{(b,c)}k_{(c,a)}}.
\end{equation}
When comparing with the physical affinities $F_1$ and $F_2$ of Eq. (\ref{eq:PhysicalUE}), we identify using Eq.~(\ref{eq:cycle2physAff}) the matrix 
\begin{equation}
  \bm{\phi} =
  \left( \begin{array}{c}
   E_b-E_a \\ E_c-E_b
  \end{array} \right).
\end{equation}
The physical currents follow from Eq.~(\ref{eq:fundamentalcylecurrent}) as $\Jmoyi{\I} =\Jmoyi{c_1}( E_b-E_a) $ and $\Jmoyi{\II}= \Jmoyi{c_1}(E_c-E_b)$, in order that 
the entropy production rate writes $\sigma= \Jmoyi{c_1}F_{c_1} = J_1 F_1 + J_2 F_2$.

We now turn to the conductance and resistance matrices. From the definition of the edge resistance matrix of Eq.~(\ref{eq:edgeResistance}), we have
$ \bar R_\exy =F_{\exy}/\Jmoyi{c_1} $ since all edge probability currents are equal to the cycle current. Eq.~(\ref{eq:CycleResistanceMatrix}) yields the cycle resistance matrix 
which is the scalar
\begin{equation}
  \label{eq:cycleconductanceUE}
  \bm{\tilde R} = \f{F_{c_1}}{\Jmoyi{c_1}}.
\end{equation}
In the end, Eq.~(\ref{eq:GegalRmoinsun}) for the non equilibrium conductance matrix yields
\begin{equation}
  \label{eq:Gijunicyclic}
  \bm{G} =\f{\Jmoyi{c_1}}{F_{c_1}}
  \left( \begin{array}{cc}
    (E_b-E_a)^2 & (E_c-E_b)(E_b-E_a) \\ (E_c-E_b)(E_b-E_a) & (E_c-E_a)^2\\
  \end{array} \right),
\end{equation}
which is not invertible as expected for unicyclic machines. In this case, the conductance matrix is equal to the minimum conductance matrix $\bm{G}_{min}$ defined in Eq.~(\ref{eq:GTC}). In the end, the parameters associated to the non-equilibrium conductance matrix are:
\begin{eqnarray}
  \label{eq:unicyclic-motor-par}
  \Pi&=(E_a-E_b)^2 (\beta_1-\beta_3)^2 \f{\Jmoyi{c_1}}{F_{c_1}} = \f{\Jmoyi{1}^2 F_{1}^{2}}{\sigma},\\
  \varphi &= \left|\f{(\beta_2-\beta_3)(E_b-E_c)}{(\beta_1-\beta_3)(E_a-E_b)}\right|= \f{-\Jmoyi{\II}F_{\II}}{\Jmoyi{\I}F_{\I}}= \eta,\\
  \xi &= -1,
\end{eqnarray}
confirming that this system operates in the tight coupling regime.

\subsection{Molecular Motor}
\label{sec:molecular-motor}

\begin{figure}[h]
  \centering
  \includegraphics[scale=0.6]{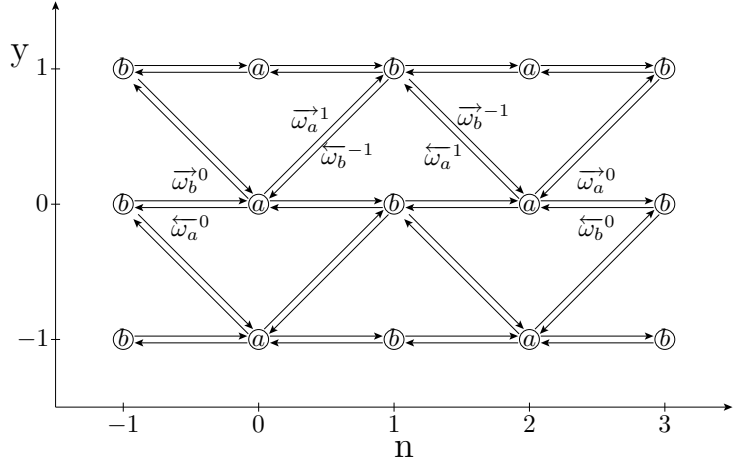}
  \caption{Sketch of the state space for our discrete model of molecular motor specifying the transition rates. The horizontal axis provides the motor position $n$ and the vertical axis the number $y$ of consumed ATP. This figure is reproduced from Ref.~\cite{Lacoste2008_vol78} with permission from authors.}
    \label{fig:sketch-mm-lattice}
\end{figure}

Our second example is a discrete model of a molecular motor \cite{Lau2007_vol99,Lacoste2008_vol78}. 
The motor has only two internal states and evolves on a linear discrete lattice by consuming Adenosine TriPhosphate (ATP) molecules.
The position of the
motor is given by two variables the position $n$ on the lattice and $y$ is the number of ATP consumed, as shown in figure~\ref{fig:sketch-mm-lattice}. The even and
odd sites are denoted by $a$ and $b$, respectively.  
Note that the lattice of $a$ and $b$ sites extends indefinitely in both directions along the $n$ and $y$ axis; for the spatial direction $n$, the lattice step defines the unit length.
There are two physical forces acting on the motor, a chemical force controlled by the chemical potential difference of the hydrolysis reaction of ATP, $\Delta \mu$ and a mechanical force $f$ applied directly on the motor. The whole system is in contact with a heat bath, and we choose to express all energies in units of $k_B T$. Equilibrium corresponds to the 
vanishing of the two currents, namely the mechanical current $\bar v$ which is the average velocity of the motor on the lattice,  
and $r$ the chemical current, which is its average rate of ATP consumption. 
Since the system operates cyclically, the change of internal energy in a cycle is zero and the first law takes the form $q+r\Delta\mu+f \bar v=0$ where $q$ is the heat flow coming from the heat bath, $r \Delta \mu$ represents the chemical work and $f \bar v$ represents the mechanical work; all quantities are evaluated in a cycle. 
Under these conditions, the second law takes the form $\sigma=- q$, and 
the entropy production rate takes the following form:
\begin{equation}
  \label{eq:MMentProd}
  \sigma =  f \bar v + r \Delta \mu.
\end{equation}
In the normal operation of the motor, chemical energy
is converted into mechanical energy, which means that the driving process $(1)$ is the chemical one and the output  
process $(2)$ the mechanical one in agreement with the choice of convention made in this paper. Thus, the two partial entropy production rates should be $\sigma_\I = r \Delta \mu$, with the chemical affinity $F_\I =  \Delta \mu$ and 
$\sigma_\II = f \bar v$, with mechanical affinity $F_\II = f$.

\begin{figure}[h]
  \centering
  \includegraphics[scale=0.6]{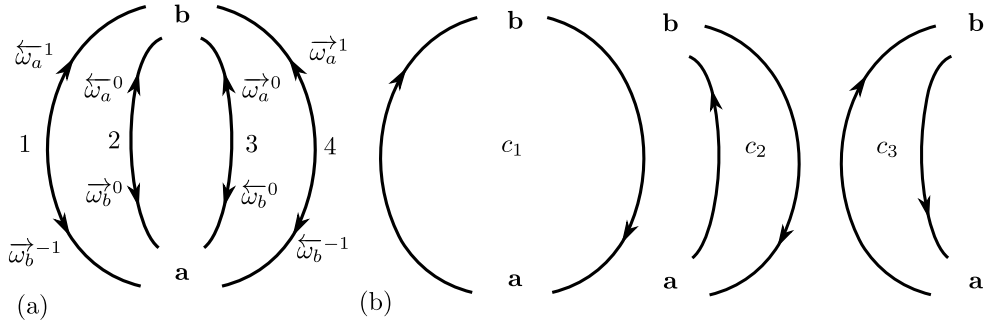}
  \caption{(a) Sketch of the effective two-state system with four edges. Edge orientation is head toward $b$. (b) Set of fundamental cycles with their orientations. \label{fig:sketch-molecular-motor}}
\end{figure}
In this model, there are four reactions between the two states, corresponding to four edges, with for each of them 
a forward or backward direction along each edge as represented in figure \ref{fig:sketch-molecular-motor}. 
Two of these reactions are passive and do not involve ATP while the other two are active and do involve ATP.
Together, there are eight rates 
for these four reactions which are given by
\begin{equation}
  \begin{array}{ll}
\overrightarrow{\omega_b}^{-1}= \alpha' e^{\theta^{+}_b f}, &\overrightarrow{\omega_b}^{0} =  \omega'\,e^{\theta^{+}_b f},\\
 \overleftarrow{\omega_a}^{1}  =  \alpha' e^{ -\epsilon + \Delta \mu - \theta^{-}_a f}, &  \overleftarrow{\omega_a}^{0} = \omega'\,e^{-\epsilon- \theta^{-}_a f},\\
  \overleftarrow{\omega_b}^{-1}= \alpha\, e^{-\theta^{-}_b f}, & \overleftarrow{\omega_b}^{0} =  \omega\,e^{-\theta^{-}_b f},  \\
 \overrightarrow{\omega_a}^{1}  =  \alpha\, e^{ -\epsilon + \Delta \mu + \theta^{+}_a f }, & \overrightarrow{\omega_a}^{0} = \omega\,e^{-\epsilon + \theta^{+}_a f} \, ,
\end{array}
\label{eq:ratesMolecularMotor}
\end{equation}
where we have kept the original notation of Refs.~\cite{Lau2007_vol99,Lacoste2008_vol78} for the rates. In the above expressions, $\theta_i^\pm$ represent load distribution factors 
 which are arbitrary except for the constraint $\theta^{+}_a+\theta^{-}_b + \theta^{-}_a+\theta^{+}_b\ =2~$\cite{Lacoste2008_vol78}.
Let us choose to orientate all these edges from state $a$ to $b$. Then, the four 
edge currents and affinities are
\begin{eqnarray}
\label{eq:MMedeLabel-1}
J_{(1)} &= \pi_a \overleftarrow{\omega_a}^{1} - \pi_b \overrightarrow{\omega_b}^{-1}, 
& \qquad F_{(1)}= \ln \frac{\overleftarrow{\omega_a}^{1}\pi_{a}}{\overrightarrow{\omega_b}^{-1}\pi_{b}}, \\
J_{(2)} &= \pi_a \overleftarrow{\omega_a}^{0} - \pi_b \overrightarrow{\omega_b}^{0}, 
& \qquad F_{(2)}= \ln \frac{\overleftarrow{\omega_a}^{0}\pi_{a}}{\overrightarrow{\omega_b}^{0}\pi_{b}}, \\
J_{(3)} &= \pi_a \overrightarrow{\omega_a}^{0} - \pi_b \overleftarrow{\omega_b}^{0}, 
& \qquad F_{(3)}= \ln \frac{\overrightarrow{\omega_a}^{0}\pi_{a}}{\overleftarrow{\omega_b}^{0}\pi_{b}}, \\
J_{(4)} &= \pi_a \overrightarrow{\omega_a}^{1} - \pi_b \overleftarrow{\omega_b}^{-1},
& \qquad F_{(4)}= \ln \frac{\overrightarrow{\omega_a}^{1}\pi_{a}}{\overleftarrow{\omega_b}^{-1}\pi_{b}},
\label{eq:MMedgeLabel-4}
\end{eqnarray}
in terms of the stationary probabilities to be in state $a$ or $b$, namely $\pi_a$ and $\pi_b$.
The explicit expressions of the currents in terms of the transition rates is known~\cite{Lau2007_vol99,Lacoste2008_vol78}.

Here, there are three cycles identified in Figure~\ref{fig:sketch-molecular-motor}c.
Given our convention of orientation of the edges, the edge currents and the cycle currents are
related in the following way
\begin{eqnarray}
J_{(1)} &=& J_{c1} + J_{c3}, \\
J_{(2)} &=& -J_{c3}, \\
J_{(3)} &=& J_{c2}, \\
J_{(4)} &=& -J_{c1} - J_{c2},
\end{eqnarray}
which means that the matrix $\bm{A}$ is
\begin{equation}
    \bm{A} = \left( \begin{array}{ccc}
    1 & 0 &1 \\
    0 & 0 & -1\\
    0& 1 &0\\
    -1&-1&0
    \end{array} \right).
\end{equation}
The physical currents can be expressed in terms of the displacements $\Delta n$ and the change in the number of ATP molecules $\Delta y$ along each transition per unit time. For the four edges, these changes are
\begin{equation}
  \label{eq:MMedegcontribution}
  \begin{array}{l|c|c|c|c}
    \text{edge}& (1) & (2) & (3) & (4)  \\
    \hline
    \Delta y & 1  & 0 & 0 & 1 \\
    \hline
    \Delta n & -1 &-1 & 1 & 1 \\
  \end{array} \qquad \quad \bm{\bar \phi} =   \left( \begin{array}{cccc}
     1 & 0 & 0 & 1 \\
    -1 &-1 & 1 & 1 \\
  \end{array} \right),
\end{equation}
which defines a matrix that we denote $\bm{\bar \phi}$.
Then, summing the edge contributions over cycles gives the matrix 
\begin{equation}
\label{matrix_phi}
  \bm{\phi}  = \bm{\bar \phi} \cdot \bm{A} =
  \left( \begin{array}{ccc}
           0 &-1&1\\
           -2 & 0 &0 \\
         \end{array} \right).
\end{equation}

Another approach is to identify the matrix $\bm{\phi}$ by making the description at the edge and cycle 
levels matches with the one at the level of physical observables. 
Precisely the entropy production rate is  
$\sigma = \sum_{i=1,4} J_{(i)} F_{(i)}$ at the edge level, $\sigma = \sum_{i=1,3} J_{c_{i}} F_{c_{i}}$,
at cycle level, and $\sigma = f \bar v + r \Delta \mu$ at the level of physical currents and affinities.
Note that all edge affinities and currents are not independent,
with the above choice of transition rates, one finds the constraint on edge currents  
$\sum_i J_{(i)}=0$ and 
similarly for edge affinities $F_{(1)}+F_{(3)}-F_{(2)}-F_{(4)}=0$.
These compatibility relations are essential to relate edge or cycle currents to the two physical currents 
$(\bar v,r)$.
The physical currents are $\bar v=2(J_{(3)}+ J_{(4)})$ and $r=-J_{(3)}-J_{(2)}$ in terms of the edge currents 
while the physical affinities are $2 f= F_{(4)}-F_{(1)}$ and $\Delta \mu= F_{(1)}-F_{(2)}$ in terms of the 
edge affinities.
In the end, the relations between the cycle affinities and the physical affinities read
\begin{eqnarray}
	F_{c_{1}} &=& - 2f, \\
	F_{c_{2}} &=& - \Delta \mu, \\
	F_{c_{3}} &=&  \Delta \mu, 
\end{eqnarray}
in agreement with the matrix $\bm{\phi}$ given in Eq.~(\ref{matrix_phi}).

 

We now turn to the conductance and resistance matrices. From the definition of the edge resistance matrix below Eq.~(\ref{eq:Dissipationboundscurrent}), we have
\begin{equation}
  \bm{\bar R} =
  \left( \begin{array}{cccc}
    \bar R_{(1)} & 0 &0 &0\\
    0&   \bar R_{(2)}& 0& 0\\
    0& 0&  \bar R_{(3)} & 0 \\
    0&0&0&   \bar R_{(4)} \\
  \end{array} \right),
\end{equation}
where using Eqs.~(\ref{eq:MMedeLabel-1}-\ref{eq:MMedgeLabel-4}), we have $\bar R_{(i)}=J_{(i)}/F_{(i)}$ for $i=1,\ldots,4$.
The cycle resistance matrix is then obtain from (\ref{eq:CycleResistanceMatrix})
\begin{equation}
  \label{eq:cycleresistancenatrixMM}
   \bm{\tilde R} =  \left( \begin{array}{ccc}
     \bar R_{(1)}+\bar R_{(4)}& \bar R_{(4)} & \bar R_{(1)} \\
     \bar R_{(4)}  & \bar R_{(3)} + \bar R_{(4)} &         0      \\
     \bar R_{(1)} &  0 &\bar R_{(1)}+ \bar R_{(2)}              
  \end{array} \right)
\end{equation}
The cycle conductance matrix is the inverse of the cycle resistance matrix (\ref{eq:cycleresistancenatrixMM}),
then this leads using (\ref{eq:defGLDF}), to the following non-equilibrium conductance matrix
\begin{equation}
  \label{eq:Gij-molecular}
\fl \bm{G} =\f{1}{Z_G}  \left( \begin{array}{cc}
  ( \bar R_{(1)}+ \bar R_{(4)} )(\bar R_{(3)}+ \bar R_{(2)}) & 2(\bar R_{(4)}   \bar R_{(2)}-\bar R_{(1)}  \bar R_{(3)}) \\
 2(\bar R_{(4)}   \bar R_{(2)}-\bar R_{(1)}  \bar R_{(3)})&  4( \bar R_{(1)}+ \bar R_{(2)} )(\bar R_{(3)}+ \bar R_{(4)}) 
  \end{array} \right), 
\end{equation}
with
\begin{equation}
  \label{eq:nomrFactormatrixGMM}
 Z_G  = \bar R_{(1)} \bar R_{(4)}  \bar R_{(3)} +  \bar R_{(1)} \bar R_{(4)}  \bar R_{(2)} +  \bar R_{(1)} \bar R_{(3)} \bar R_{(2)} + \bar R_{(4)}  \bar R_{(3)} \bar R_{(2)}.
\end{equation}
Using this matrix and the definitions of Eq.~(\ref{eq:new-parameters}), we find the following parameters
\begin{eqnarray}
  \label{eq:molecularmotor-motor-par}
  \Pi &=& \f{( \bar R_{(1)}+ \bar R_{(4)} )(\bar R_{(3)}+ \bar R_{(2)})}{ \bar R_{(1)} \bar R_{(4)}  \bar R_{(3)} +  \bar R_{(1)} \bar R_{(4)}  \bar R_{(2)} +  \bar R_{(1)} \bar R_{(3)} \bar R_{(2)} + \bar R_{(4)}  \bar R_{(3)} \bar R_{(2)}}(\Delta\mu)^2,\\
  \varphi &=&\sqrt{\f{ ( \bar R_{(1)}+ \bar R_{(2)} )(\bar R_{(3)}+ \bar R_{(4)})}{( \bar R_{(1)}+ \bar R_{(4)} )(\bar R_{(3)}+ \bar R_{(2)})}}\f{2|f|}{|\Delta \mu|} ,\\
  \xi&=& \f{ -(\bar R_{(4)}   \bar R_{(2)}-\bar R_{(1)}  \bar R_{(3)}) }{\sqrt{( \bar R_{(1)}+ \bar R_{(2)} )(\bar R_{(3)}+ \bar R_{(4)}) ( \bar R_{(1)}+ \bar R_{(4)} )(\bar R_{(3)}+ \bar R_{(2)})}},
\end{eqnarray}
which are used to make the plots of  Figs.~\ref{fig:etaxi} and \ref{Fig:Power-efficiency}.

\subsection{Discussion}

\begin{figure}
  \centering
  \includegraphics[width=14cm]{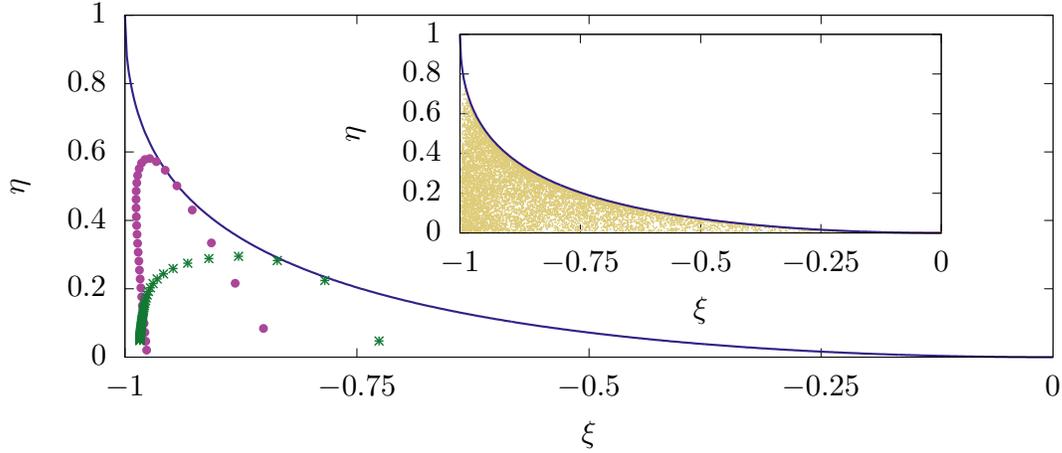}
  \caption{Illustration of the bound of Eq.~(\ref{eq:Max-Eff}) (blue solid line) for the molecular motor model. For a given chemical potential difference $\Delta\mu=15.0$ (purple circles), the force $f$ is varied along the curve. Alternatively, for a given force $f=-1$ (green crosses) the chemical potential difference is varied along the curve. The kinetic parameters are those of Ref.~\cite{Lacoste2008_vol78}: $\alpha=0.57$, $\alpha'=1.3.10^{-6}$, $\omega=3.5$, $\omega'=108.15$ $\epsilon=10.81$, $\theta_a^+=0.25$, $\theta_a^-=1.83$, $\theta_b^+=0.08$, $\theta_b^-=-0.16$. Inset: Efficiency versus degree of coupling when varying all kinetic parameters at fixed affinities $\Delta \mu =10.0$ and $f=-1.9$. The kinetic parameters listed above are randomly chosen  
by multiplying the values used in the main figure by $e^{x}$ with $x$ drawn uniformly within $[-2,2]$.}
  \label{fig:etaxi}
\end{figure}

The maximal efficiency given by Eq.~(\ref{eq:Max-Eff}) is shown as function of the degree of coupling in Fig.~(\ref{fig:etaxi}).   
This maximal efficiency is compared with the efficiency of the molecular motor model which is analytically solvable. The 
corresponding comparison for the unicyclic engine is not shown because the degree of coupling is always $-1$.
In order to test this bound, we vary either (i) the thermodynamic forces, namely $f$ and $\Delta \mu$, which together characterize the distance to equilibrium, or (ii) the kinetic parameters of the model ($\alpha$, $\alpha'$, $\epsilon$, $\theta_i$..). 
The test (i) is carried out in the main figure in which either the affinity $f$ is varied at fixed $\Delta \mu$ or vice versa, 
covering a large regime of conditions far from equilibrium. The test (ii) is carried out in the inset, by scanning over a large panel of kinetic parameters.  
Both figures confirm that the maximum efficiency only depends on the degree of coupling. These figures also show that this maximum
efficiency is reached under some accessible conditions.

\begin{figure}
\centering
\includegraphics[width=14cm]{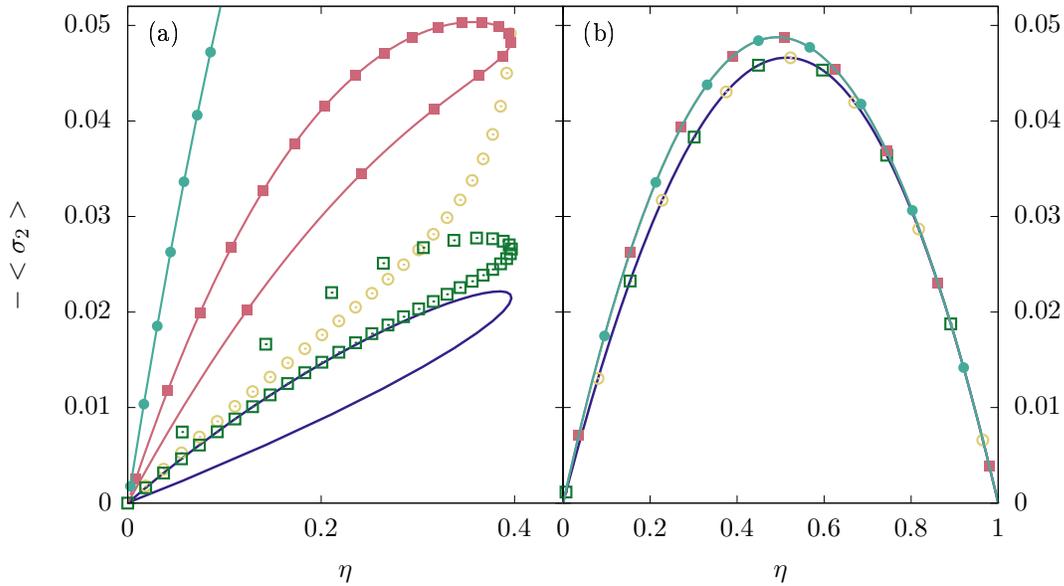}
\caption{Output entropy production rate as a function of the machine efficiency 
using exact expression (solid blue line). Power-efficiency bounds of
 Eqs.~(\ref{eq:PowerEffS1}) (green open squares),~(\ref{eq:PowerEffS2}) 
(yellow open circles),~(\ref{eq:PowerEffTradeOffVariance}) (magenta full 
squares),~(\ref{eq:PowerEffTradeOffVariance2}) (blue full circles) for (a) 
the molecular motor model with $\Delta\mu=7.0$ and the 
same kinetic parameters as in Fig.~\ref{fig:etaxi} and (b) the unicyclic engine
 at $\beta_1=0.5$, $\beta_3=1$, $\Gamma=1$, $E_a=1$,
 $E_b=4$ and $E_c=2$. For both figures, $F_\I$ is held fixed when varying 
$F_2$. \label{Fig:Power-efficiency}
 }
\end{figure}

Figures \ref{Fig:Power-efficiency}a and \ref{Fig:Power-efficiency}b illustrate the power-efficiency trade-off for the molecular motor and the unicyclic engine respectively, by showing the mean output entropy production rate $-\langle \sigma_2 \rangle$ as function of the efficiency $\eta$.
A striking feature in these plots is that the entropy production rate is bi-valued for the molecular motor as explained in section \ref{sec:power-eff}
while it is single-valued for the unicyclic engine, because the unicyclic engine is a tight coupled engine. 
In order to test the inequality of Eqs.~(\ref{eq:PowerEffS1}--\ref{eq:PowerEffS2}), we compare 
 $- \langle \sigma_2 \rangle$ (solid line) evaluated using exact expressions of the average currents, with the power-efficiency bounds of  Eqs.~(\ref{eq:PowerEffS1}--\ref{eq:PowerEffS2}) (empty symbols).
As shown in Fig.~\ref{Fig:Power-efficiency}b, these bounds become exact in the tight coupled case.

The figure also shows a comparison with the power-efficiency inequalities derived by Pietzonka and Seifert \cite{Pietzonka2017_vol}
 (full symbols). 
The variances appearing in these inequalities can be evaluated from the cumulant generating function of the currents
 that is known exactly for these models \cite{Pietzonka2016_vol93,Lau2007_vol99,Lacoste2008_vol78}.
We confirm with this figure that the new bound derived from the present framework is more tight 
than the bounds derived in Ref.~\cite{Pietzonka2017_vol}, in agreement with Eqs.~(\ref{eq:combine}). Note that the two bounds derived in this reference collapse with each other in the tight coupled case, but 
stay above the exact value except at the two extremal values of $\eta=0$ and $\eta=1$. Indeed, in these regions, 
the engine works near equilibrium.
One reason for which the bounds 
of Ref.~\cite{Pietzonka2017_vol} are less tight than ours is that they are a consequence of uncertainty
relations, which for this molecular motor model do not yield  
a particularly sharp prediction of the efficiency, 
as discussed in Ref. \cite{Baiesi2017}.

\section{Conclusion}
\label{sec:conclusion}

In this work, we have developed a framework based on the notion of non-equilibrium conductance
matrix to analyze the efficiency and the output power of energy converters 
operating far from equilibrium. 
This matrix is initially only partially constrained by the dependence of the physical currents on thermodynamic affinities.
Nevertheless it shares many properties with the Onsager matrix, the two matrices are 
symmetric positive definite and become identical near equilibrium but differ otherwise. 
These properties are sufficient to exploit a parametrization of the efficiency introduced by Kedem and Kaplan for 
machines operating near equilibrium~\cite{Kedem1965_vol61} and use it for general machines operating far from equilibrium.
With this parametrization  linked to a specific choice of 
non-equilibrium conductance matrix, we have shown that the efficiency of 
 machines is generally bounded by an 
 universal expression dependent only on the degree of coupling. 
The maximum value of this bound is the reversible efficiency which is only 
accessible to tight coupled machines.
This result means practically that a bound on the efficiency of a machine 
can be deduced from a measurement of its degree of coupling. 
This observation could have interesting applications for various thermodynamic devices, 
such as for instance photoelectric cells. 
 
When a microscopic kinetic model of the machine is known, more insights   
into the efficiency of the machine can be obtained.
In particular, tighter bounds on the output power in terms of the efficiency 
as compared to Ref.~\cite{Pietzonka2017_vol} 
follow from our approach. 
We have explained the relation between the various 
bounds using properties on the large deviations of the currents. 

This work naturally begs the question whether the non-equilibrium conductance 
matrix can itself be determined experimentally. 
As mentioned above the dependence of the physical currents on thermodynamic affinities is 
in general insufficient to define the conductance matrix uniquely.
However we have also shown that a unique conductance matrix can be defined from the 
knowledge of local resistances (which make up the resistance matrix) and 
of the weights between cycle currents and physical currents (which make up the
 $\bm{\phi}$ matrix).  

The framework introduced here should be useful to revisit old questions 
such as the efficiency at maximum power or the role played of time reversal symmetry for 
the efficiency. 
In this context, it would be interesting to study extensions of the present framework   
to systems in which the Onsager reciprocity relations are modified, either
due to a magnetic field \cite{Benenti2011_vol106,Bonella2017_vol96} 
or because the machine operates under time-periodic driving \cite{Proesmans2015_vol115}.

\section*{Aknowlegements}
We acknowledge H.-J. Hilhorst for his pertinent comments on this paper.

\appendix

\section{Microscopic framework for the non-equilibrium conductance matrix}
\label{MicroToMacro}

In the following, we emphasize the physical meaning of $\bm{G}$ as a conductance matrix. We intend to  shown how to switch from the resistance matrix at the edge level to the NE conductance matrix at the level of physical currents.

Starting at the edge level, the resistance matrix $\bm{\bar R}$ is diagonal in the space of edges, with diagonal elements $\bar R _{\exy}$ as defined in Eq.~(\ref{eq:edgeResistance}).
The inverse of the edge resistance matrix is the edge conductance matrix $\bm{\bar G} = \bm{\bar R}^{-1}$. We remark that the elements of the resistance matrix depend on the physical affinities through transition rates and stationary probability.  

At the level of cycles, the matrix $\bm{\tilde R}$ for cycle resistance introduced in Eq.~(\ref{eq:CycleResistanceMatrix}) connects cycle affinities and currents via
\begin{equation} \label{eq:cycle-resistance}
	  F_{c} = \sum_{c'} \tilde R_{c,c'} \Jmoyi{c}.
\end{equation}
Indeed, using Eq.~(\ref{eq:cyclecurrent2edge}) in  Eq.~(\ref{eq:edgeResistance}), one may express $\bm{\tilde{R}}$ as a function of $\bm{\bar R}$ since
\begin{eqnarray}
 F_\exy &=& \sum_{c} \bar R_{\exy}A_{\exy,c} \Jmoyi{c} , \\
    F_{c'}& =&  \sum_{c}  \sum_{\exy} ({A}^\mathrm{T})_{c',\exy} {\bar R}_{\exy}{A}_{\exy,c} \Jmoyi{c}    ,
\end{eqnarray}
where we have used Eq.~(\ref{eq:edge2cycleAff}) in the second step.
This leads to the cycle resistance matrix defined in the main text
\begin{equation} \label{eq:expression-cycle-resistance}
	\tilde R_{c',c}  = \sum_{\exy}(A^\mathrm{T})_{c,\exy} \bar R_{\exy} A_{\exy,c}. 
      \end{equation}
Here the analogy with electric circuits holds: electrical resistances add when connected in series. 
The cycle conductance matrix $\bm{\tilde G} $ is then 
\begin{equation}
	\bm{\tilde G}  \equiv \bm{\tilde R}^{-1} = ( \bm{A}^\mathrm{T}\cdot \bm{\bar R}\cdot \bm{A} )^{-1}= \bm{A}^{+} \cdot \bm{\bar G }\cdot \bm{A}^{\mathrm{T}+},
\end{equation}
where $\bm{A}^{+}$ is the Moore-Penrose pseudo inverse of the matrix of fundamental cycles $\bm{A}$ \cite{Book_BenIsrael2003}.

At the level of physical observables, the NE conductance connects currents to affinities via
\begin{equation}
\Jmoyi{\Y} \equiv \sum_\X {G}_{\Y,\X}  F_{\X}.
\end{equation}
Considering that the amount of physical quantity $\Y$ exchanged with the environment during cycle $c$ is $\phi_{\Y,c}$ and using Eqs.~(\ref{eq:cycle-resistance}), one gets 
\begin{equation}
		\Jmoyi{\Y} = \sum_c\phi_{\Y,c} \Jmoyi{c} =  \sum_{c,c'} {\phi}_{{\Y},c} {\tilde G}_{c,c'} F_c  = \sum_\X  \sum_{c,c'} {\phi}_{{\Y},c} {\tilde G}_{c,c'} {\phi}_{c,{\X}}^\mathrm{T} F_{{\X}}.
              \end{equation}
Therefore, the physical conductance matrix writes
\begin{equation}
    \label{eq:NonEqConducmatrixcomplet}
	\bm{G} = \bm{\phi} \cdot \bm{A}^{+}  \cdot \bm{\bar G}  \cdot  ( \bm{\phi}\cdot \bm{A}^{+} )^\mathrm{T},
\end{equation}
which is the same non-equilibrium matrix as given by Eqs.~(\ref{eq:defGLDF}--\ref{eq:GegalRmoinsun}). The electrical analogy also holds:  cycle conductances add when connected in parallel which makes senses when considering that the current flows from one reservoir to another through sequences of cycles.

\section*{References}

\bibliographystyle{unsrt}
\bibliography{Base_papier}

\end{document}